\documentclass[10pt]{iopart}

\usepackage{iopams,epsfig,citesort}
\eqnobysec
\begin{document}

\title[Disorder and spin dynamics in f-electron NFL systems]{Disorder, inhomogeneity and spin dynamics in f-electron non-Fermi liquid systems}

\author{\boldmath D~E MacLaughlin\dag, R~H Heffner\ddag, O~O Bernal\S, K Ishida$\mathbf{\|}$, J~E Sonier\P, G~J Nieuwenhuys$\mathbf{^{+}}$, M~B Maple$\sharp$ and G~R Stewart\dag\dag}

\address{\dag\ Department of Physics, University of California, Riverside, California 92521, USA}

\address{\ddag\ Los Alamos National Laboratory, Los Alamos, New Mexico 87545, USA}

\address{\S\ Department of Physics and Astronomy, California State University, Los Angeles, California 90032, USA}

\address{$\|$\ Department of Physics, Graduate School of Science, Kyoto University, Kyoto 606-8502, Japan}

\address{\P\ Department of Physics, Simon Fraser University, Burnaby, British Columbia, Canada V5A 1S6}

\address{$^{+}$\ Kamerlingh Onnes Laboratory, Leiden University, 5400 RA Leiden, The Netherlands}

\address{$\sharp$\ Physics Department and IPAPS, University of California, San Diego, La Jolla, California 92093, USA}

\address{\dag\dag\ Department of Physics, University of Florida, Gainesville, Florida 32611, USA}

\begin{abstract}
Muon spin rotation and relaxation ($\mu$SR) experiments have yielded evidence that structural disorder is an important factor in many f-electron-based non-Fermi-liquid (NFL) systems. Disorder-driven mechanisms for NFL behaviour are suggested by the observed broad and strongly temperature-dependent $\mu$SR (and NMR) linewidths in several NFL compounds and alloys. Local disorder-driven theories (Kondo disorder, Griffiths-McCoy singularity) are, however, not capable of describing the time-field scaling seen in muon spin relaxation experiments, which suggest cooperative and critical spin fluctuations rather than a distribution of local fluctuation rates. A strong empirical correlation is established between electronic disorder and slow spin fluctuations in NFL materials.\\[12pt] (\today)
\end{abstract}



\maketitle

\section{Introduction}

The observed breakdown of Landau's Fermi-liquid paradigm in a number of metallic systems has led to an explosion of effort to understand this {\em non-Fermi liquid\/} (NFL) behaviour~\cite{ITP96iop,Stew01,VNvS02}. The NFL breakdown is usually identified by deviations of low-temperature thermodynamic and transport properties [specific heat~$C(T)$, magnetic susceptibility~$\chi(T)$, electrical resistivity~$\rho(T)$] from the predictions of Landau Fermi-liquid theory [$C(T) \propto T$, $\chi(T) = \rm const.$, $\Delta\rho(T) = \rho(T) - \rho(0) \propto T^2$]~\cite{Nozi64}. Such deviations been observed in many f-electron heavy-fermion metals and alloys~\cite{Stew01}, and often (but not always) take specific forms: the specific heat Sommerfeld coefficient $\gamma(T) = C(T)/T$ diverges logarithmically as $T \to 0$; $\chi(T)$ varies as $1 - T^{1/2}$ or $-\ln T$ or $T^{-m}$ ($m \approx 1/3$); and $\Delta\rho(T)$ varies as $T^n$, $n < 2$.

By definition no phase transition occurs at a nonzero temperature in an ideal NFL system. Nevertheless, in most NFL systems a magnetic phase is nearby in the appropriate phase diagram; the transition temperature~$T_{\rm c}(\delta)$ is reduced by varying an experimental parameter~$\delta$ (pressure, composition, magnetic field, \dots), and NFL behaviour is observed in the neighborhood of the critical parameter value~$\delta_{\rm c}$ for which $T_{\rm c}$ is just suppressed to zero. This is a {\em quantum critical point\/} (QCP)~\cite{Sach99}, which separates magnetically ordered and paramagnetic phases at zero temperature. Most theoretical effort in this area has concentrated on effects of the QCP\@. Magnetic resonance and other measurements have shown, however, that {\em structural disorder\/} is an important component of the NFL phenomenon in many systems.

Nuclear magnetic resonance (NMR) and transverse-field muon spin rotation (TF-$\mu$SR, i.e., magnetic field applied transverse to the muon spin) measurements in UCu$_4$Pd and UCu$_{3.5}$Pd$_{1.5}$~\cite{BMLA95,BMAF96,MBL96} demonstrated that the magnetic susceptibility in these NFL materials is strongly inhomogeneous. Following this discovery two broad classes of theories began to address the role of disorder in NFL behaviour. In the `Kondo-disorder' approach~\cite{BMLA95,MDK96,MDK97,MDK97a} 
structural disorder gives rise to a broad distribution of Kondo temperatures~$T_{\rm K}$\@. NFL behaviour arises from low-$T_{\rm K}$ spins that are not in the Fermi-liquid state. In the `Griffiths-McCoy singularity' picture~\cite{CNCJ98,CNJ00} spin-spin interactions freeze low-$T_{\rm K}$ f moments into clusters with a wide distribution of sizes; the larger clusters dominate the susceptibility and lead to divergent behaviour as the temperature is lowered. Both Kondo-disorder and Griffiths-McCoy singularity scenarios seem to be compatible with observed NMR and TF-$\mu$SR linewidths~\cite{LMCN00,YMRI04}.

Longitudinal-field muon spin relaxation (LF-$\mu$SR) experiments in NFL UCu$_{5-x}$Pd$_x$, $x = 1.0$ and 1.5~\cite{MBHN01,MHBN02}, and CePtSi$_{1-x}$Ge$_x$, $x = 0$ and 0.1~\cite{MRYB03}, indicate, however, that the f-ion spin dynamics in these alloys are better described by a picture in which critical slowing down occurs cooperatively throughout the sample, rather than at (rare) low-$T_{\rm K}$ spins or large clusters. The muon spin relaxation rates are nevertheless widely distributed, and the dynamic behaviour closely resembles that of spin glasses~\cite{KMCL96,KBCL01}. Thus an important question is whether 
disorder in the interactions dominates the dynamics, as in a spin glass above the glass temperature, or if, instead, the critical slowing down is still controlled by a QCP\@. Theoretical treatments exist~\cite{GrRo99,GPS00} that combine elements of both these viewpoints, and describe a `quantum spin glass' with a suppressed or vanishing glass temperature.

In the meantime NFL systems have been discovered in which disorder does not appear to play an essential role. At ambient pressure these include CeCu$_{5.9}$Au$_{0.1}$~\cite{LPPS94,AFGS95,Loeh96}, Ce(Ru$_{1-x}$Rh$_{x}$)$_2$Si$_2$, $x \approx 0.5$~\cite{YMTK99,YMTO99}, CeNi$_2$Ge$_2$~\cite{JPGM96,SBGL96}, and YbRh$_2$Si$_2$~\cite{TGML00}. Thus one can compare spin dynamics in `disordered' and `ordered' materials. The result of this comparison is evidence that the rapid muon spin relaxation in the disordered systems is due to slow dynamics associated with quantum spin-glass behaviour, rather than (homogeneous) quantum criticality. This evidence consists of 
\begin{itemize}
\item the observation of a wide inhomogeneous distribution of relaxation rates characteristic of disordered systems when the relaxation is strong,
\item the fact that in the ordered NFL compounds~CeNi$_2$Ge$_2$ and YbRh$_2$Si$_2$, and even in the doped alloys~CeCu$_{5.9}$Au$_{0.1}$ and Ce(Ru$_{0.5}$Rh$_{0.5}$)$_2$Si$_2$, the low-frequency weight of the spin fluctuation spectrum is more than an order of magnitude smaller than in the disordered NFL systems, and
\item remarkably good correlation of the muon spin relaxation rate and its inhomogeneity with the residual resistivity of NFL systems.
\end{itemize}

This paper reviews $\mu$SR studies of NFL materials that shed light on the interplay between disorder and NFL behaviour. Candidate disorder-driven NFL mechanisms are discussed in \sref{sec:ddNFL}, which describes the way in which TF-$\mu$SR (and NMR) linewidths probe the spatial inhomogeneity of the magnetic susceptibility that is the principal signature of disorder-driven NFL behaviour. As an example, TF-$\mu$SR evidence for a disorder-driven NFL mechanism in the NFL compound~UCu$_4$Pd is described in \sref{sec:UCu4Pd}. \Sref{sec:dynamics} 
presents evidence that muon spin relaxation functions in at least some of these materials obey {\em time-field scaling\/}~\cite{KMCL96,KBMC00,KBCL01}: the dependence of the spatially-averaged muon spin relaxation function~$\overline{G}(t,H)$ on time~$t$ and longitudinal field $H$ obeys the scaling relation~\cite{KMCL96}
\begin{equation}
\overline{G}(t,H) = \overline{G}(t/H^\gamma) \,.
\label{eq:scaling}
\end{equation}
Time-field scaling is a signature of slow dynamics, i.e., a zero-frequency singularity in the fluctuation noise spectrum. It can in principle appear near any critical point, but is usually associated with spin-glass-like behaviour. In NFL materials such glassy spin dynamics may arise from the effect of disorder on quantum critical fluctuations. In \sref{sec:correl} muon spin relaxation results are compared with the residual electrical resistivity (a measure of disorder in the electronic system) and the low-temperature Sommerfeld coefficient (a measure of the characteristic electronic energy scale), and found to correlate well with the former but not the latter. These results establish strong links between quantum criticality, slow fluctuations, and electronic disorder in NFL materials~\cite{MRYB03,MacL03}. The $\mu$SR experiments and questions they raise raised are summarized in \sref{sec:discus}.

\section{Disorder-driven NFL mechanisms\label{sec:ddNFL}}

As described briefly in the Introduction, NMR and TF-$\mu$SR experiments~\cite{BMLA95,MBL96,BMAF96} have revealed anomalously large and strongly temperature-dependent resonance linewidths in a number of NFL materials. It was realized~\cite{BMLA95} that this behaviour is characteristic of a spatially inhomogeneous distribution of local f-electron susceptibilities. This in turn suggests that a characteristic excitation energy~$\Delta$ associated with the f electrons, which enters the zero-temperature susceptibility via the relation~$\chi \propto 1/\Delta$, is inhomogeneously distributed due to structural disorder.

\begin{figure}[ht]
\flushright \includegraphics[clip=,width=4in]{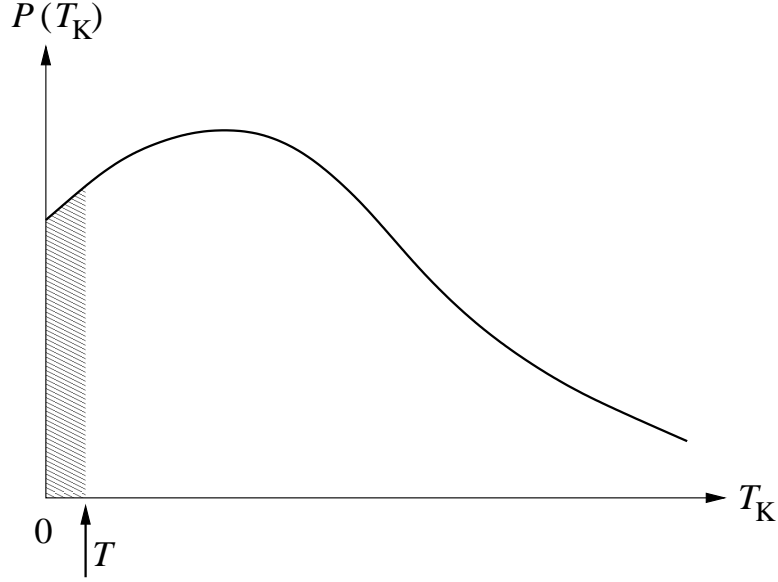}
\caption{Distribution function~$P(T_{\rm K})$ in the Kondo-disorder model. At a given temperature~$T$ impurities with $T_{\rm K} < T$ (shaded region) are not compensated and give rise to the non-Fermi-liquid behaviour of the system. This behaviour continues down to $T = 0$ if $P(T_{\rm K}{=}0)$ is nonzero. After reference \protect\cite{MacL00}.}
\label{fig:tkdist}
\end{figure}

In the single-ion `Kondo-disorder' picture~\cite{BMLA95,MDK96} this energy is the Kondo temperature~$T_{\rm K}$, which is broadly distributed because of its extreme sensitivity to f-electron--conduction-electron hybridization. If the distribution function~$P(T_{\rm K})$ is broad enough so that $P(T_{\rm K}{=}0) \ne 0$, as shown in \fref{fig:tkdist}, 
then at any nonzero temperature~$T$ the f ions with $T_{\rm K} < T$ are not Kondo compensated 
and are the source of NFL behaviour in the Kondo-disorder model. Bulk thermodynamic and transport properties are averages over $P(T_{\rm K})$ of the usual Kondo expressions for these quantities. It can be shown~\cite{MDK96} that $\gamma(T) \propto \chi(T) \propto -\ln T$, and $\Delta\rho(T) \propto T$.

A simple heuristic calculation may clarify how the inhomogeneous spread $\delta\chi$ in $\chi$ might be expected to behave in a Kondo-disorder model. Assume a spread~$\delta T_{\rm K}$ in Kondo temperatures~$T_{\rm K}$. Assume further that the Curie-Weiss expression
\begin{equation}
\chi(T) \propto (T + T_{\rm K})^{-1}\,,
\label{eq:CW}
\end{equation}
captures the essence of the local Kondo physics. Then $\delta\chi$ is given by
\begin{eqnarray}
\delta\chi(T) & \approx \left| \frac{\partial\chi}{\partial T_{\rm K}} \right| \delta T_{\rm K}  \\
&\propto (T + T_{\rm K})^{-2}\, \delta T_{\rm K}\propto \chi^2(T)
\end{eqnarray}
for small $\delta T_{\rm K}$. Thus $\delta\chi/\chi$ should be roughly proportional to $\chi$, with temperature an implicit parameter, reflecting the increasing effect of the distribution in $T_{\rm K}$ in equation \eref{eq:CW} as the temperature is lowered. We do not expect this picture to be very accurate, as it assumes a small spread in $T_{\rm K}$ and a simplified Curie-Weiss susceptibility, but a strong increase in $\delta\chi/\chi$ at large $\chi$ (low temperatures) 
is evidence for the basic effect.

The Griffiths-McCoy singularity model~\cite{CNCJ98,CNJ00} also invokes a broad distribution of $T_{\rm K}$, but assumes that uncompensated f-ion spins interact via RKKY coupling and form clusters. These clusters then tunnel between orientational configurations via Kondo spin-flip processes, and the distributed energy~$\Delta$ characterizes these tunneling processes. The model exhibits Griffiths-McCoy singularities associated with a QCP below percolation threshold. The thermodynamic properties diverge as $T \to 0$ in a manner described by a nonuniversal exponent~$\lambda < 1$. Predictions for the specific heat~$C(T)$, the magnetic susceptibility~$\chi(T)$, and the fractional rms width~$\delta\chi(T)/\chi(T)$ of the inhomogeneous susceptibility distribution are given by
\begin{equation}
\gamma(T) \propto \chi(T) \propto T^{-1+\lambda}\,; 
\end{equation}
\begin{equation}
\delta\chi(T)/\chi(T) \propto T^{-\lambda/2} \,.
\end{equation}
Thermodynamic properties have been shown to be in good agreement with this theory for a number of NFL alloy systems~\cite{dACDD98}.

For both the Kondo-disorder and Griffiths-McCoy singularity models fits to the bulk susceptibility determine the distribution function~$P(\chi)$, from which $\delta\chi(T)/\chi(T)$ can be calculated. This can then be compared with NMR or TF-$\mu$SR linewidth data with no further adjustable parameters~\cite{BMLA95,MBL96,BMAF96,CNCJ98} as described below; good agreement in a given NFL material can be taken as evidence for applicability of a disorder-driven mechanism of this kind. This procedure often works equally well for both the Kondo-disorder and Griffiths-McCoy singularity scenarios~\cite{LMCN00,YMRI04}, so that other experiments are necessary to distinguish between them.

\subsection{NMR/TF-$\mu$SR linewidth and susceptibility inhomogeneity} \label{sec:deltachi}

We briefly review the relation between the 
static resonance linewidth and the spread in local susceptibility. More complete descriptions are given in references~\cite{MBL96} and~\cite{MacL00}.

In a paramagnet in an applied magnetic field~$\bi{B}_0$ the time-average or static local field~$\langle \bi{B}_{\rm L}\rangle$ at the site of a {\em spin probe\/} (muon or nucleus) is shifted by an amount~$\Delta \bi{B} = \langle\bi{B}_{\rm L}\rangle - \bi{B}_0$ due to the hyperfine coupling (direct or transferred) between the spin probe and the surrounding electronic spin polarization. The relative shift~$K_i = |\Delta \bi{B}|_i/|\bi{B}_0|$ of the $i^{\rm th}$ spin probe is given by
\begin{equation}~K_i = \sum_j a_{ij}\chi_j\,, \label{eq:Kchi} \end{equation}
where $\chi_j$ is the susceptibility of the $j^{\rm th}$ f ion and $a_{ij}$ is the transferred hyperfine coupling constant between spin probe~$i$ and ion~$j$. A spatially inhomogeneous distribution of the $\chi_j$ gives rise to a corresponding distribution of the $K_i$, and hence broadens the resonance line.

From equation \eref{eq:Kchi} the relation between the rms width~$\delta\chi$ of the susceptibility distribution and the rms width~$\delta K$ of the shift distribution can be shown to be 
\begin{equation} 
\delta K = a^{\textstyle *}\delta\chi\,, 
\label{eq:KD}
\end{equation}
where $a^{\textstyle *}$ is an effective hyperfine coupling constant that depends on the spatial correlation of the disordered susceptibility and can be calculated from the $a_{ij}$~\cite{MBL96}. Thus the spread in shifts is a direct measure of the spread in $\chi$, a quantity that is very hard to obtain from other experiments. Determination of $a^{\textstyle *}$ in general requires knowledge of the spatial correlation function that describes disorder in the inhomogeneous susceptibility; this correlation function can be determined from spin-probe spectra only if data are available for more than one spin probe~\cite{MBL96,BMAF96}. The calculation simplifies in the extreme limits of long-range correlation (LRC) (correlation length $\gg$ range of the hyperfine coupling) and short-range correlation (SRC) (correlation length $\lesssim$ nearest-neighbor f-ion distance). It is found that
\begin{equation}
a^{\textstyle *}_{\rm LRC} = \left|\mbox{$\sum_j$} a_{ij}\right| \quad {\rm and} \quad a^{\textstyle *}_{\rm SRC} = \left(\mbox{$\sum_j$} 
a_{ij}^2\right)^{\!\!1/2} \,.
\end{equation}
Thus one can find $\delta\chi$ from $\delta K$. Furthermore, the quantity~$\delta K/(a^{\textstyle *}\chi)$ is an estimator of the fractional spread in susceptibility~$\delta\chi/\chi$.

A convenient way to compare the data to these model results is to plot $\delta K/(a^{\textstyle*}\chi)$ [cf.\ equation \eref{eq:KD}] and the theoretical $\delta\chi(T)/\chi(T)$ (cf.\ \sref{sec:ddNFL}) versus $\chi(T)$, with temperature an implicit parameter~\cite{BMLA95,MBL96}. An example of such a comparison is given below for the NFL compound UCu$_4$Pd.

\subsection{Evidence for a disorder-driven NFL mechanism in UCu\/$_4$Pd \rm \protect\cite{MHNL98}} \label{sec:UCu4Pd}

Heavy-fermion materials in the series~UCu$_{5-x}$Pd$_x$, $1 \lesssim x \lesssim 1.5$, exhibit NFL behaviour~\cite{AnSt93,ChMa96}. These alloys crystallize in the fcc AuBe$_5$ structure (space group $F\overline{4}3m$). The end compound~UCu$_5$ possesses two crystallographically inequivalent copper sites in the ratio $4 : 1$ at the $16e$ and $4c$ positions (Wyckoff notation), which are filled by Pd ions in the alloys. Therefore stoichiometric UCu$_4$Pd could crystallize in an ordered structure, with only $4c$ sites occupied by Pd ions, in which case disorder might not play a role in the NFL behaviour of this compound.

Chau \etal~\cite{CMR98} reported elastic neutron diffraction measurements on members of the UCu$_{5-x}$Pd$_x$ series, and found no evidence for Pd/Cu disorder for $x = 1$. But the neutron scattering cross sections for Pd and Cu are similar, and the diffraction data do not rule out the possibility of interchange between Pd and Cu sites at the level of $\sim$4\% occupation of $16e$ (Cu) sites by Pd atoms [and therefore $\sim$16\% occupation of $4c$ (Pd) sites by Cu atoms]. 

\begin{figure}[ht]
\flushright \includegraphics[clip=,width=4in]{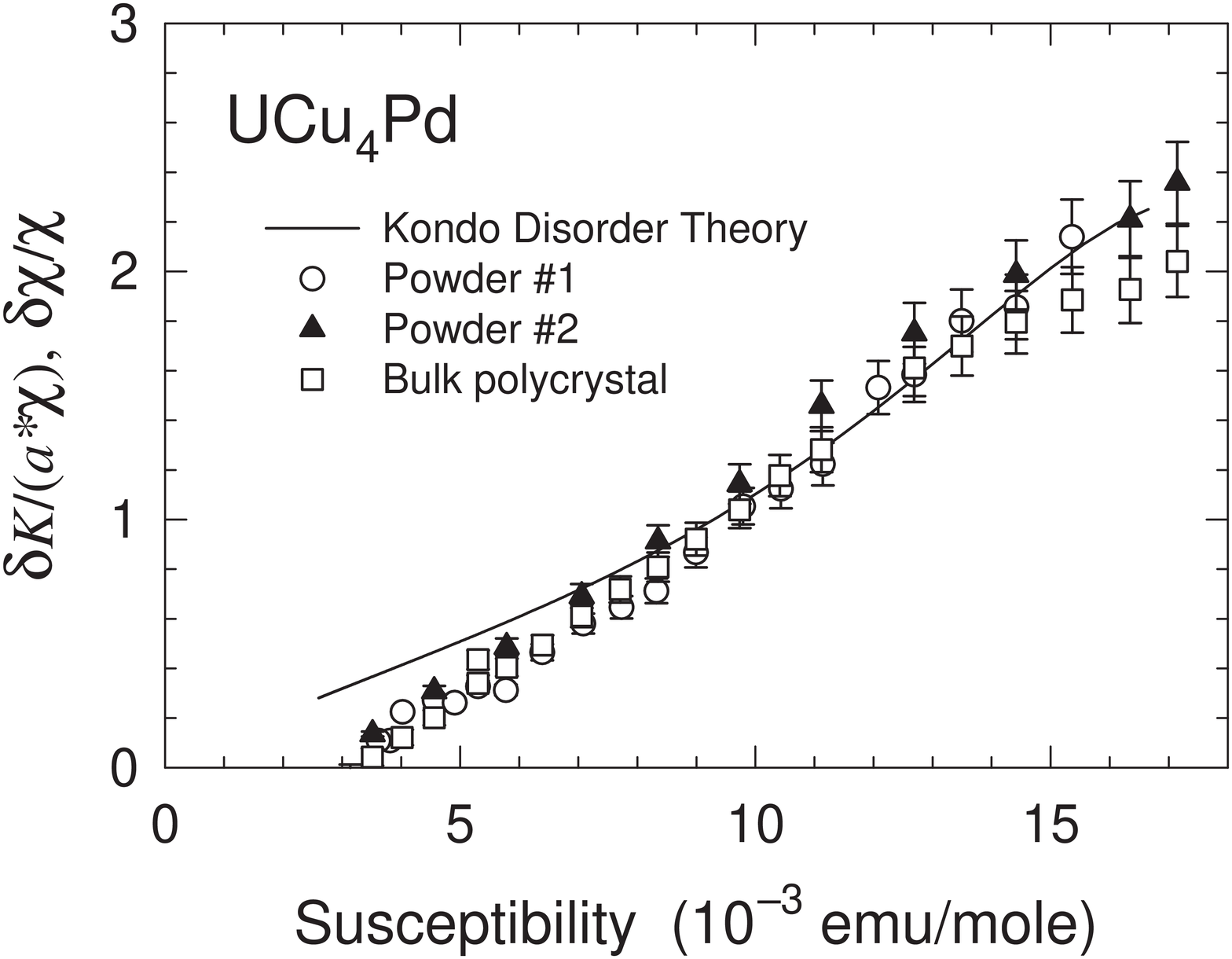}
\caption{Dependence of $\delta K/(a^{\textstyle*}\chi)$ on bulk susceptibility~$\chi$, with temperature an implicit parameter, in UCu$_4$Pd. Powder~\#1: previously-studied powder sample~reference \protect\cite{BMAF96}. Powder~\#2: powder sample used in neutron diffraction studies~reference \protect\cite{CMR98}. Curve: $\delta\chi/\chi$ from the Kondo-disorder theory~reference \protect\cite{BMLA95,BMAF96,MBL96}. From reference \protect\cite{MacL00}.} 
\label{fig:dchionchi}
\end{figure}

TF-$\mu$SR linewidths have been measured in a number of samples of UCu$_4$Pd~\cite{MHNL98}, including a previously-studied powder sample~\cite{BMAF96} (Powder~\#1), the powder sample used for the neutron diffraction experiments (Powder~\#2), and a bulk polycrystal consisting of a few single crystals. \Fref{fig:dchionchi} 
plots $\delta K/(a^{\textstyle*}\chi)$, with $a^{\textstyle*}$ calculated in the SRC limit, vs.\ the bulk susceptibility~$\chi$, with temperature an implicit parameter~\cite{MacL00}. (A combination of TF-$\mu$SR and NMR results has shown that the SRC limit is appropriate to this system~\cite{BMAF96}.) The data exhibit the near proportionality to $\chi$ expected from the simple heuristic argument given above. (The dropoff for $\chi \lesssim 5 \times 10^{-3}$~emu/mole (i.e., for $T \gtrsim 100$~K) is not well understood but could be due to thermally-activated muon diffusion, which would motionally average the TF-$\mu$SR line and reduce its width.) The samples were all prepared differently and are likely to have a range of defect concentrations, in which case the susceptibility inhomogeneity is remarkably insensitive to the amount of structural disorder. We note that the low-temperature values of $\delta K/(a^{\textstyle*}\chi)$ in UCu$_4$Pd and UCu$_{3.5}$Pd$_{1.5}$ (which is definitely a disordered alloy) are also very similar~\cite{BMAF96}. Shown for comparison in \fref{fig:dchionchi} is $\delta\chi/\chi$ from the Kondo-disorder model~\cite{BMLA95,BMAF96,MBL96}. The disorder-driven Griffiths-McCoy singularity theory (not shown) is also in good agreement with the data~\cite{CNCJ98}. 

X-ray absorption fine structure (XAFS) studies in UCu$_4$Pd have confirmed that Cu/Pd site interchange is substantial. Initially the observed disorder was argued to be sufficient to produce the required broad distribution of $T_{\rm K}$ in the single-ion Kondo disorder model~\cite{BMHC98}, but subsequent annealing studies~\cite{BBKC02,BSKW02} call this into question. The XAFS results nevertheless leave little doubt that there is considerable interplay between site-exchange disorder and NFL behaviour in UCu$_4$Pd.

\section{Muon spin relaxation and NFL spin dynamics\label{sec:dynamics}}

\subsection{Time-field scaling\label{sec:timefield}}

We have seen that disorder-driven mechanisms have been considered for the NFL properties of some f-electron systems~\cite{MDK96,CNCJ98}, and it is natural to consider the possibility of extremely disordered or `glassy' behaviour in the spin dynamics of these systems. On theoretical and experimental grounds it is believed that glassy dynamics lead to long-time correlations with distinct signatures as the freezing or `glass' temperature~$T_g$ is approached from above~\cite{KMCL96}. In a spin glass the spin autocorrelation function~$q(t) = \langle \bi{S}_i(t)${\boldmath $\cdot$}$\bi{S}_i(0) \rangle$ is theoretically predicted to exhibit power-law ($q(t) = ct^{-\alpha}$) or `stretched-exponential' ($q(t) = c\exp[-(\lambda t)^\beta]$) behaviour~\cite{PSAA84}. Power-law correlation has been found in spin-glass {\em Ag\/}Mn using muon spin relaxation~\cite{KMCL96}. 

In a magnetic field~$H$ applied parallel to the muon spin direction muon spin relaxation (LF-$\mu$SR) is due to thermally-excited f-electron spin fluctuations that couple to the muons. A muon at a given site experiences a time-varying local field~${\bi{B}}_{\rm loc}(t)$ due to fluctuations of neighboring f moments. Following Keren \etal\cite{KMCL96,Kere04}, under motionally narrowed conditions the local muon asymmetry~$G(t,H)$ relaxes exponentially:
\begin{equation}
G(t,H) = \exp\left[ -2\Delta^2 \tau_{\rm c}(H)t \right] \,,
\label{eq:relax}
\end{equation}
where $\Delta^2 = \gamma_\mu^2\langle |{\bi{B}}_{\rm loc}|^2\rangle$ is the time-averaged mean-square coupling constant in frequency units, $\gamma_\mu$ is the muon gyromagnetic ratio, and the local correlation time~$\tau_{\rm c}(H)$ is given by
\begin{equation}
\tau_{\rm c}(H) = \int_0^\infty\!\! dt\, q(t) \cos(\omega_\mu t) = cu_{\rm c}(H) \,;
\label{eq:tauc}
\end{equation}
here $\omega_\mu = \gamma_\mu H$ is the muon Zeeman frequency. We consider $\Delta$ and the prefactor~$c$ but not the functional form of $u_{\rm c}(H)$ to vary from site to site in a disordered material. Then the sample-averaged asymmetry~$\overline{G}(t,H)$ is given by
\begin{equation}
\overline{G}(t,H) = \int\!\!\!\int d\Delta\,dc\, \rho(\Delta,c) \exp\left[ -2\Delta^2 c\,u_{\rm c}(H)t \right] \,,
\label{eq:Gav}
\end{equation}
where $\rho(\Delta,c)$ is the joint distribution function for $\Delta$ and $c$. A plot of $\ln\overline{G}(t,H)$ vs.\ $t$ always exhibits upward curvature, because the initial slope on such a plot is the spatial average of the rate whereas the long-time slope is dominated by regions with low rates. Such upward curvature is difficult to achieve by other means, and is therefore evidence for spatial inhomogeneity in $c$ and/or $\Delta$.

From equation \eref{eq:Gav} the time and field dependence of $\ln\overline{G}(t,H)$ enter in the combination~$u_{\rm c}(H)t$. This means that $\overline{G}(t,H)$ scales as this combination independently of the form of $\rho(\Delta,c)$. For both the power-law and stretched-exponential forms of $q(t)$ 
\begin{equation}
u_{\rm c}(H) \propto H^{-\gamma}
\label{eq:scaluc}
\end{equation}
from equation \eref{eq:tauc} [in the high-field limit for stretched-exponential $q(t)$]~\cite{KMCL96}. Then $u_{\rm c}(H)t$ scaling of $\overline{G}(t,H)$  results in the time-field scaling relation~$\overline{G}(t,H) = \overline{G}(t/H^\gamma)$, i.e., a plot of $\overline{G}(t,H)$ versus $t/H^\gamma$ is universal for the correct choice of $\gamma$. The sign of $\gamma - 1$ distinguishes between power-law ($\gamma < 1$) and stretched-exponential ($\gamma > 1$) correlations~\cite{KMCL96}. It is important to note that no particular form for $\overline{G}(t)$ is assumed in this analysis.

As noted above, the early- and late-time slopes of $\ln\overline{G}(t,H)$ are determined by spatially different sets of muons. If $\overline{G}(t)$ obeys time-field scaling, the shape of $\overline{G}(t)$ is independent of field, which implies that $\tau_{\rm c}(H)$ [cf.\ equation \eref{eq:tauc}] has the same functional dependence for all sites in the sample. Thus observation of time-field scaling is evidence that the {\em local\/} dynamics exhibit scaling, with an exponent that is the same at all sites even if the sample is disordered. The significance of this property is discussed in \sref{sec:local?}.

\subsection{UCu\/$_{5-x}$Pd\/$_x$, $x = 1.0$ and \rm 1.5\label{sec:ucupd}}

\subsubsection{Muon spin relaxation \rm \protect\cite{MBHN01}.\label{sec:muonrelax}}

Evidence from LF-$\mu$SR experiments suggests that spin dynamics in the NFL alloys~UCu$_{5-x}$Pd$_x$, $x = 1.0$ and 1.5, are {\em glassy\/}, i.e., strongly slowed by disorder~\cite{Mydo93}. The sample average muon spin relaxation function~$\overline{G}(t,H)$ (i.e., the asymmetry in muon counting rate; see, e.g., \cite{Sche85}) is strongly sub-exponential, indicating a quenched inhomogeneous distribution of relaxation rates as discussed above, and obeys the time-field scaling relation~$\overline{G}(t,H) = \overline{G}(t/H^\gamma)$ for applied magnetic field~$H$ between $\sim$15~Oe and $\sim$1~kOe. The field dependence corresponds to a measurement of the Fourier transform of $q(t)$ over the frequency range~$\gamma_\mu H/2\pi \approx 200$~kHz--14~MHz, where $\gamma_\mu = 2\pi \times 13.55$~kHz/Oe is the muon gyromagnetic ratio. Power-law behaviour of $q(t)$ is implied by the observation~$\gamma < 1$~\cite{KMCL96}, and also by the temperature-frequency scaling found in the inelastic neutron scattering (INS) cross section~\cite{AORL95a} (the LF-$\mu$SR and INS results are compared in more detail in \sref{sec:tempdep}). The LF-$\mu$SR time-field scaling measurements extend the frequency range over which power-law correlations are observed in UCu$_{5-x}$Pd$_x$ downward by three orders of magnitude. In addition, zero-field $\mu$SR (ZF-$\mu$SR) shows no sign of static magnetism or spin freezing in UCu$_4$Pd above 0.05~K\@. This together with glassy scaling points to some form of {\em quantum\/} spin-glass behaviour~\cite{Sach98,GrRo99}. 

\begin{figure}[ht]
\flushright \includegraphics[clip=,width=4in]{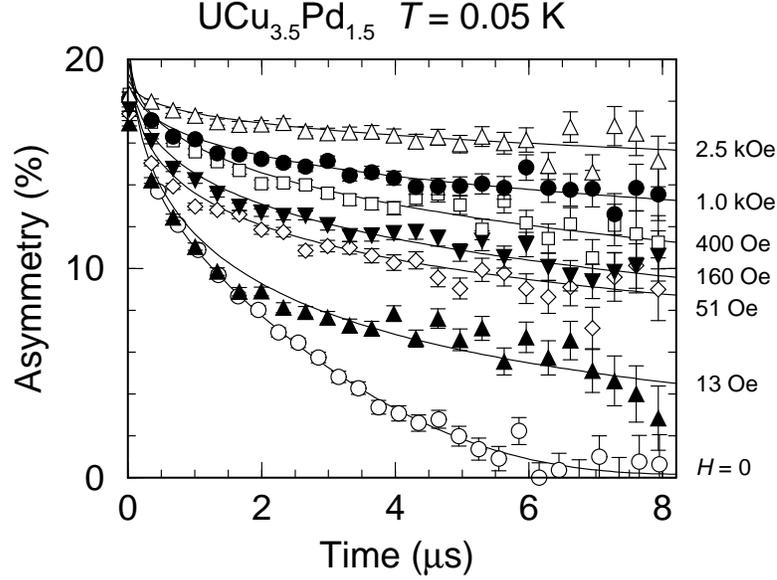}
\caption{Field dependence of sample-averaged muon asymmetry relaxation function~$\overline{G}(t)$ in UCu$_{3.5}$Pd$_{1.5}$, $T = 0.05$~K\@. Curves: fits as described in text. From reference \protect\cite{MBHN01}.}
\label{fig:UC3Pdasy}
\end{figure}

\Fref{fig:UC3Pdasy} 
gives $\overline{G}(t)$ in UCu$_{3.5}$Pd$_{1.5}$ for $T = 0.05$~K and values of applied field~$H$ between 0 and 2.5~kOe.\footnote[1]{Figures 3--6 are reprinted with permission from D E MacLaughlin, O O Bernal, R H Heffner, G J Nieuwenhuys, M S Rose, J E Sonier, B Andraka, R Chau, and M B Maple, Phys.\ Rev.\ Lett.\ 87, 066502 (2001). Copyright \copyright\ (2001) by the American Physical Society.} The relaxation slows with increasing field. For low enough fields we expect the field dependence to be due to the change of $\omega_\mu$ [cf.\ equations \eref{eq:relax}--\eref{eq:Gav}] rather than a direct effect of field on $q(t)$; a breakdown of scaling would occur for high fields where this might cease to be true. 

\begin{figure}[ht]
\flushright \includegraphics[clip=,width=4in]{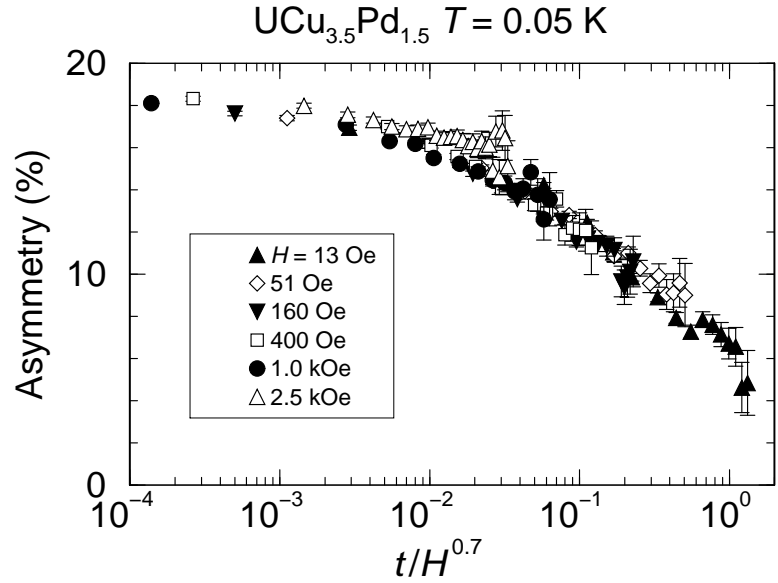}
\caption{Data of \protect\fref{fig:UC3Pdasy} plotted versus the scaling variable~$t/H^{0.7}$. From reference \protect\cite{MBHN01}.} 
\label{fig:UC3Pscal}
\end{figure}

\begin{figure}[ht]
\flushright \includegraphics[clip=,width=4in]{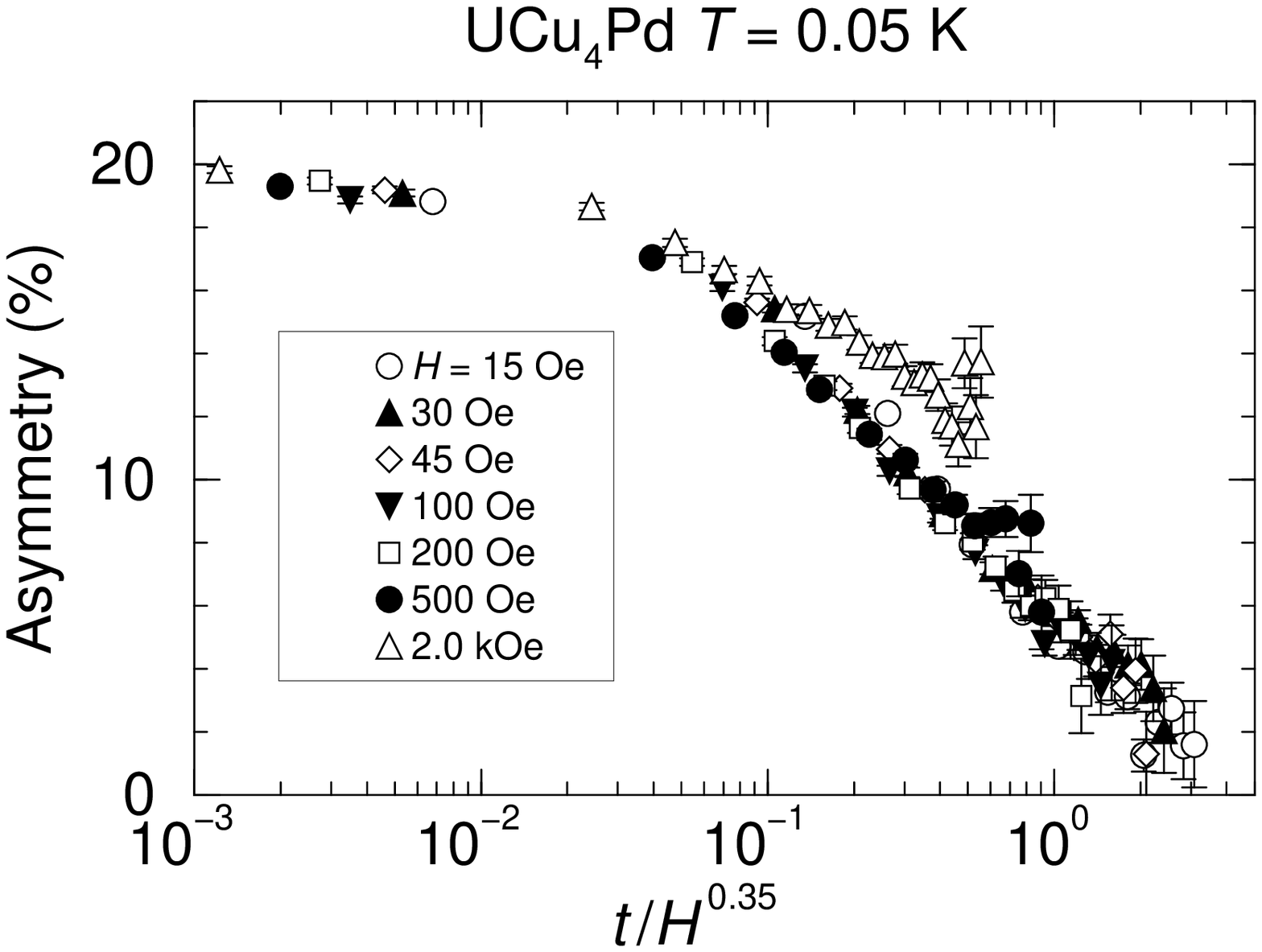}
\caption{Scaling plot of $\overline{G}(t,H)$ vs.\ $t/H^{0.35}$ for UCu$_4$Pd, $T = 0.05$~K\@. From reference \protect\cite{MBHN01}.} 
\label{fig:UC4Pscal}
\end{figure}

The same asymmetry data are plotted in \fref{fig:UC3Pscal} 
as a function of the scaling variable~$t/H^\gamma$. For $\gamma = 0.7 \pm 0.1$ the data scale well over more than three orders of magnitude in $t/H^\gamma$ and for all fields except 2.5~kOe. Fields $\mu_{\rm B}H \gtrsim k_{\rm B}T$ would be expected to affect the spin dynamics, and indeed the static susceptibility of UCu$_4$Pd is suppressed by fields $\sim$1~kOe below $\sim$0.5~K (Vollmer \etal \cite{SSHK98}). The scaling exponent~$\gamma$ is less than 1, implying that $q(t)$ is well approximated by a power law (or a cutoff power law~\cite{KMCL96}) rather than a stretched-exponential or exponential. From the $\mu$SR data $q(t) \approx ct^{-0.3 \pm 0.1}$. We note again that no specific form for the muon asymmetry function has been assumed. A scaling plot is given in \fref{fig:UC4Pscal} 
for UCu$_4$Pd, $T = 0.05$~K\@. Here the scaling exponent~$\gamma = 0.35 \pm 0.1$ is significantly smaller than in UCu$_{3.5}$Pd$_{1.5}$. Scaling again breaks down for high enough fields.

The fluctuation-dissipation theorem\footnote{The fluctuation-dissipation theorem may be invalid in the frozen spin-glass state because of broken ergodicity, but presumably holds in the high-temperature state considered here. See, e.g., Fischer K H and Hertz J A 1991 {\em Spin Glasses\/} (Cambridge: University Press) p~138.} relates $\tau_{\rm c}(H)$ to the imaginary component~$\chi''(\omega)$ of the local ($q$-independent) f-electron dynamic susceptibility:
\begin{equation}
\tau_{\rm c}(H) \approx \frac{k_{\rm B}T}{\mu_{\rm B}^2} \left( \frac{\chi''(\omega)}{\omega} \right) 
\label{eq:FDthm}
\end{equation}
for $\hbar\omega \ll k_{\rm B}T$\@. INS experiments above $\sim$10 K~\cite{AORL95a,AORL95b} yield
\begin{equation}
\overline{\chi''}(\omega,T)$ as $\omega^{-\gamma} F(\hbar\omega/k_{\rm B} T) \,,
\label{eq:chi}
\end{equation}
with $\gamma = 0.33$ and $F(x) = \tanh(x/1.2)$ for both UCu$_4$Pd and UCu$_{3.5}$Pd$_{1.5}$. Using this form in the limit~$x \ll 1$, $\tau_{\rm c}(H)$ obtained from equation \eref{eq:FDthm} is independent of $T$ and proportional to $H^{-\gamma}$; the latter is in accord with the LF-$\mu$SR time-field scaling. The INS value of $\gamma$ agrees with LF-$\mu$SR data at $T = 0.05$~K for UCu$_4$Pd ($\gamma = 0.35$), but not for UCu$_{3.5}$Pd$_{1.5}$ ($\gamma = 0.7$) (but see \sref{sec:tempdep}).

To go further one must fit the LF-$\mu$SR data to an appropriate functional form for the asymmetry. On purely empirical grounds we use the stretched-exponential
\begin{equation}
\overline{G}(t) = \exp[-(\Lambda t)^K]\,. 
\label{eq:stretch}
\end{equation}
Values less than unity of the `stretching power'~$K$ (not to be confused with the Knight shift) yield sub-exponential relaxation corresponding to a distribution of relaxation rates~(see above and, e.g., reference~\cite{KMCL96}). This form is used because it characterizes an {\em a priori\/} unknown relaxation-rate distribution, and because $\Lambda$ conforms with the general definition of the relaxation time~$1/\Lambda$ as the time at which $\overline{G}(t)$ decays to $1/\rm e$ of its initial value. For $H = 0$ (ZF-$\mu$SR) the data were fit to the product of equation \eref{eq:stretch} and the zero-field Kubo-Toyabe (K-T) function~\cite{HUIN79} characteristic of static relaxation by nuclear dipolar fields at muon sites. This form is expected when the muon local field has both static nuclear dipolar and dynamic f-moment contributions. A nuclear dipolar field $\sim$2.3~Oe was measured in both alloys for $T \gg 1$~K, where the contribution of U-moment fluctuations to the zero-field muon spin relaxation rate becomes negligible. Nonzero values of $H$ were chosen large enough to decouple the muon spin relaxation from the nuclear dipolar field~\cite{HUIN79} leaving only the dynamic U-moment contribution, so relaxation data for these fields were fit to equation \eref{eq:stretch} without the K-T function. The curves in \fref{fig:UC3Pdasy} are examples of these fits.

\begin{figure}[ht]
\flushright \includegraphics[clip=,width=4in]{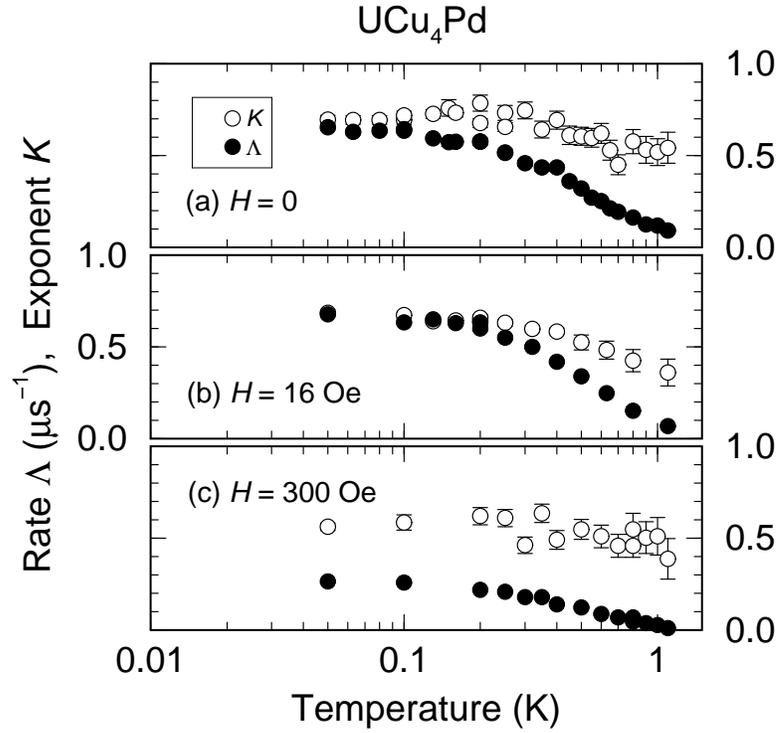}
\caption{Temperature dependence of muon stretched-exponential relaxation rate~$\Lambda$ (filled circles) and power~$K$ (open circles) in UCu$_4$Pd. ($a$)~Applied longitudinal field~$H = 0$. ($b$)~$H = 16$~Oe. ($c$)~$H = 300$~Oe. Note that $\Lambda$ and $K$ do not have the same dimensions; they are plotted on the same graph only for convenience. From reference \protect\cite{MBHN01}.}
\label{fig:UCu4Pd_temp}
\end{figure}

\Fref{fig:UCu4Pd_temp}
 gives $\Lambda(T)$ and $K(T)$ for UCu$_4$Pd at three values of $H$\@. With decreasing temperature $\Lambda$ increases slowly and saturates to a constant below 0.11--0.2~K\@. As noted previously the INS scaling predicts a temperature-independent relaxation rate,   in mild disagreement with the observed weak temperature dependence of $\Lambda(T)$. The stretching power~$K$ is 0.4--0.5 at 1~K, indicative of a broad distribution of relaxation rates~\cite{KMCL96}, and increases slightly with decreasing temperature. Similar behaviour is exhibited by $\Lambda(T)$ and $K(T)$ in UCu$_{3.5}$Pd$_{1.5}$ (data not shown), with rates slower than in UCu$_4$Pd by $\sim$30\% at low fields and $\sim$100\% at 100--300~Oe due to the larger scaling exponent~$\gamma$.

No anomaly is found in the ZF-$\mu$SR data at temperatures $\sim$0.1--0.2~K, where specific heat and ac susceptibility measurements (in different samples) suggest spin-glass-like freezing~\cite{SSHK98,VPvLC00}. The muon--f-moment coupling in UCu$_{5-x}$Pd$_x$ is predominantly dipolar~\cite{BMAF96} with a coupling field~$0.55 \pm 0.05~{\rm kOe}/\mu_{\rm B}$. Randomly-frozen moments of the order of 1~$\mu_{\rm B}$/U ion would result in a ZF-$\mu$SR relaxation rate ${\sim}50~\mu{\rm s}^{-1}$, two orders of magnitude larger than the observed rate. This result places an upper bound of ${\sim}10^{-3}~\mu_{\rm B}$/U ion on any frozen moment in UCu$_4$Pd or UCu$_{3.5}$Pd$_{1.5}$. The discrepancy with the results of \cite{SSHK98} might arise from differences in annealing conditions, or from a difference in sensitivity of the two techniques to superparamagnetic clusters or inclusions~\cite{Mydo93}. Such clusters could contribute significantly to the magnetic susceptibility, but might not occupy enough sample volume to be observed in ZF-$\mu$SR experiments.

\subsubsection{Scaling exponent: temperature dependence \rm \protect\cite{MHBN02}.\label{sec:tempdep}}

\begin{figure}[ht]
\flushright \includegraphics[clip=,width=4in]{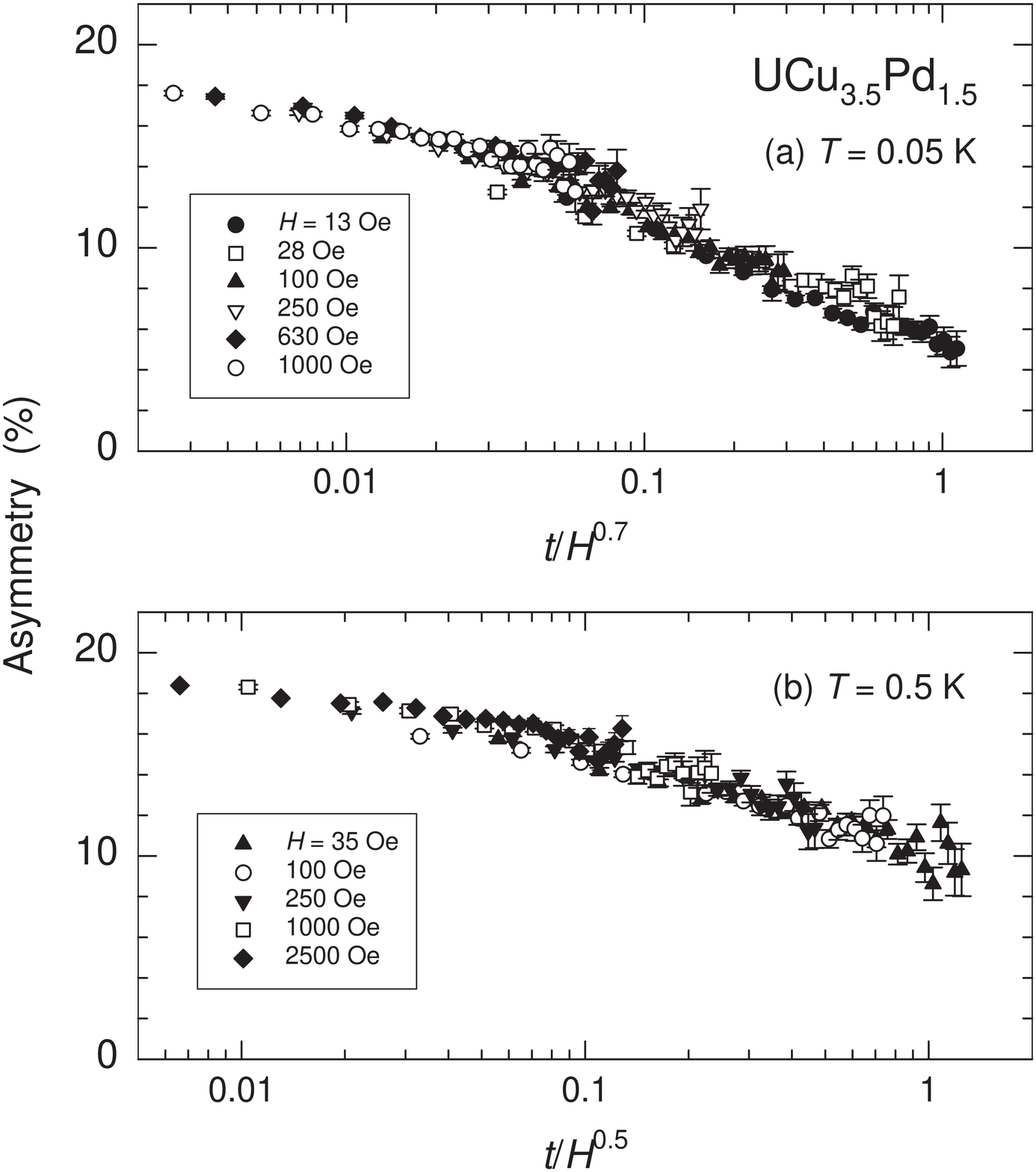}
\caption{Dependence of sample-average muon asymmetry relaxation function~$\overline{G}(t)$ on scaling variable~$t/H^\gamma$ in UCu$_{3.5}$Pd$_{1.5}$. ($a$)~$T = 0.05$~K, $\gamma = 0.7\pm0.1$. ($b$)~$T = 0.5$~K, $\gamma = 0.5\pm0.1$. From reference \protect\cite{MHBN02}.}
\label{fig:UC35Pscaling}
\end{figure}

\begin{figure}[ht]
\flushright \includegraphics[clip=,width=4in]{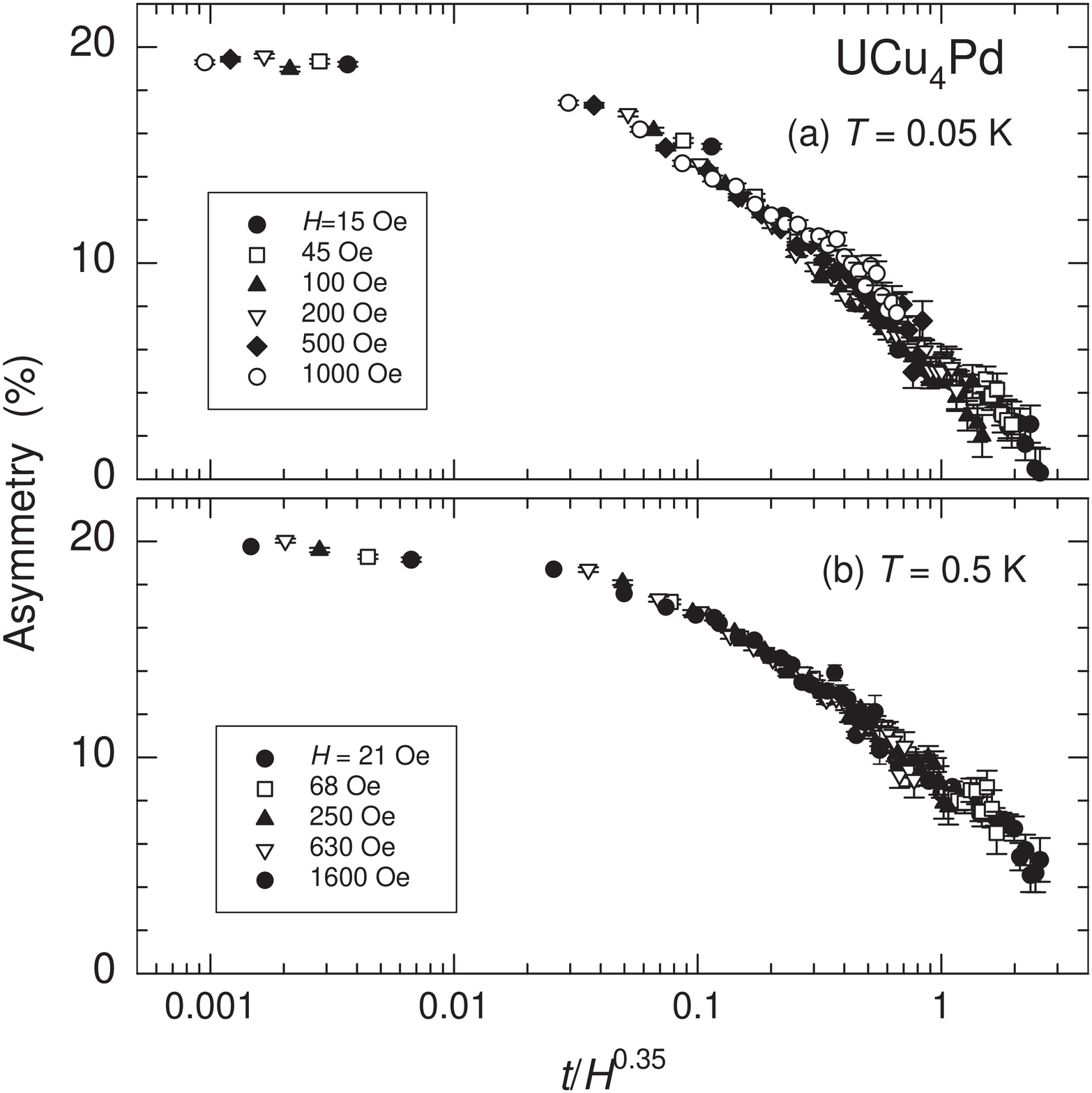}
\caption{Dependence of $\overline{G}(t)$ on $t/H^\gamma$ in UCu$_4$Pd. ($a$)~$T = 0.05$~K\@. ($b$)~$T = 0.5$~K\@. At both temperatures $\gamma = 0.35 \pm 0.10$. From reference \protect\cite{MHBN02}.}
\label{fig:UC4Pscaling}
\end{figure}

Time-field scaled LF-$\mu$SR asymmetry data for UCu$_{3.5}$Pd$_{1.5}$ and UCu$_4$Pd are given in figures~\ref{fig:UC35Pscaling} and \ref{fig:UC4Pscaling}, 
respectively, for temperatures of 0.05~K and 0.5~K\footnote{Figures 7--8 are reprinted from Physica B, vol~312-313, MacLaughlin D~E, Heffner R~H, Bernal O~O, Nieuwenhuys G~J, Sonier J~E, Rose M~S, Chau R, Maple M~B and Andraka B, ``Slow spin dynamics in non-{Fermi}-liquid {UCu$_{5-x}$Pd$_x$}, $x = 1.0$ and 1.5,'' pp~453--455, Copyright \copyright\ (2002), with permission from Elsevier.}\@. In UCu$_{3.5}$Pd$_{1.5}$ $\gamma$ from muon spin relaxation exhibits a temperature dependence, becoming smaller at higher temperatures: $\gamma(0.05~{\rm K}) = 0.7\pm0.1,\ \gamma(0.5~{\rm K}) = 0.5\pm0.1$. This behaviour is reminiscent of spin-glass {\em Ag\/}Mn above the spin-freezing `glass' temperature~$T_g$, where $\gamma$ increases as $T \rightarrow T_g$~\cite{KMCL96}. It suggests a monotonic temperature dependence of $\gamma$ in UCu$_{3.5}$Pd$_{1.5}$ between the low-temperature muon spin relaxation results and the higher-temperature (${\gtrsim}10$~K) INS value~$\gamma = 0.33$~\cite{AORL95b}. 

In UCu$_4$Pd, on the other hand, $\gamma = 0.35\pm0.10$ from muon spin relaxation at both 0.05~K and 0.5~K (\fref{fig:UC4Pscaling}). This value is 
in agreement with INS experiments above 10~K~\cite{AORL95b}, which give $\gamma = 0.33$ as in UCu$_{3.5}$Pd$_{1.5}$. The temperature independence of $\gamma$ in UCu$_4$Pd suggests that the slow fluctuations are quantum rather than thermal in origin. In UCu$_{3.5}$Pd$_{1.5}$ the Pd concentration~$x$ is closer to the value $x \approx 2$ above which a spin-glass phase is observed~\cite{AnSt93}, and the proximity of this phase may be reflected in the low-temperature increase of $\gamma$.

\subsection{CePtSi\/$_{1-x}$Ge\/$_x$ \rm \protect\cite{MRYB03}}

CePtSi$_{1-x}$Ge$_x$ is a NFL system~\cite{SGGO94} with similarities to CeCu$_{6-x}$Au$_{x}$, except that the residual resistivity of the latter alloy is somewhat lower. The end compound CePtSi is a heavy-fermion paramagnet. Ge doping expands the lattice and favors antiferromagnetism. A nonzero N\'eel temperature~$T_N$ appears for $x \gtrsim 0.1$, which is therefore a candidate for a QCP\@. But magnetic susceptibility and $^{29}$Si NMR studies of CePtSi and CePtSi$_{0.9}$Ge$_{0.1}$~\cite{YMRI04} show that both alloys exhibit strong disorder in the susceptibility above 2~K, which when extrapolated to lower temperatures accounts for much of the NFL specific heat. 

\begin{figure}[ht]
\flushright \includegraphics[clip=,width=4in]{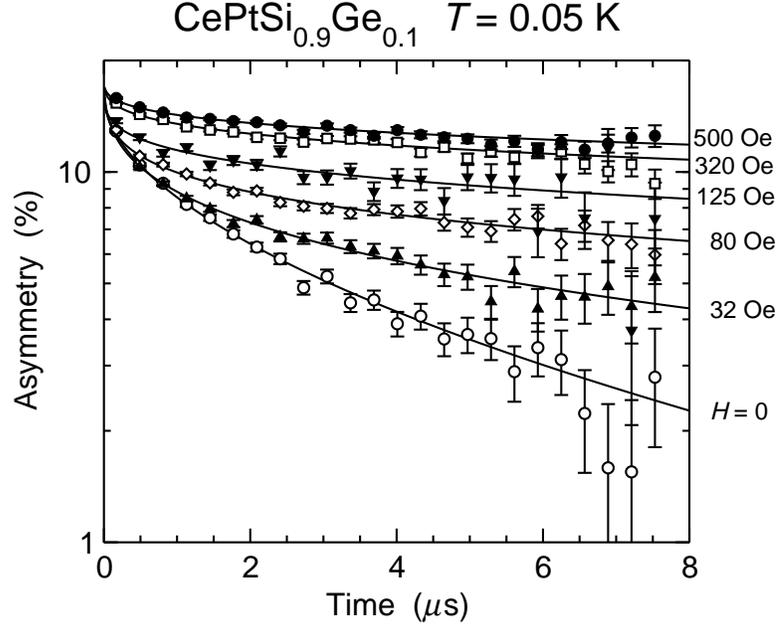}
\caption{Semi-log plots of LF-$\mu$SR asymmetry data in CePtSi$_{0.9}$Ge$_{0.1}$, $T = 0.05$~K\@. Curves: fits to stretched exponential $A\exp[-(\Lambda t)^K]$. From reference \protect\cite{MRYB03}.}
\label{fig:NFLfig1}
\end{figure}

\begin{figure}[ht]
\flushright \includegraphics[clip=,width=4in]{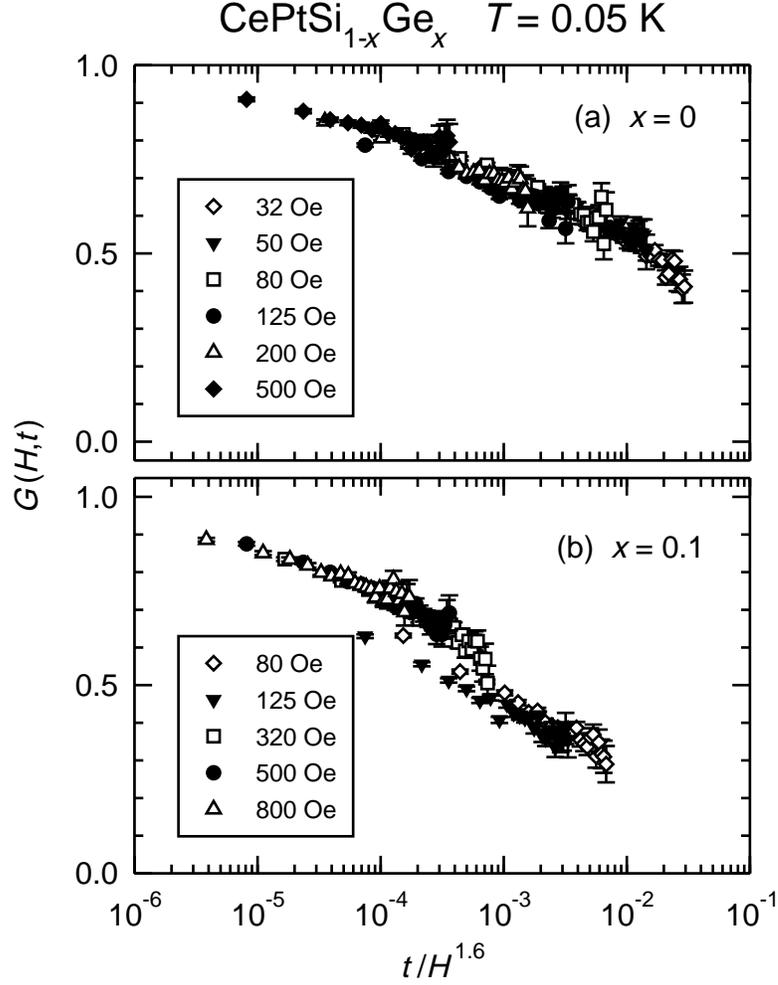}
\caption{Dependence of LF-$\mu$SR relaxation function~$G(t,H)$ on scaling variable~$t/H^{1.6}$ in CePtSi$_{1-x}$Ge$_x$, $T = 0.05$~K\@. ($a$)~$x = 0$. ($b$) $x = 0.1$. From reference \protect\cite{MRYB03}.}
\label{fig:NFLfig2}
\end{figure}

LF-$\mu$SR experiments below 1~K in CePtSi$_{1-x}$Ge$_x$, $x = 0$ and 0.1, exhibit rapid relaxation and time-field scaling behaviour qualitatively similar to that seen in UCu$_{5-x}$Pd$_x$, $x = 1.0$ and 1.5~\cite{MRYB03}. \Fref{fig:NFLfig1} 
gives the muon decay asymmetry relaxation function for CePtSi$_{0.9}$Ge$_{0.1}$ at $T = 0.05$~K and several field values. Normalized relaxation functions~$G(t,H)$ for CePtSi$_{1-x}$Ge$_x$ at $T = 0.05$~K are plotted in \fref{fig:NFLfig2} 
against the scaling variable~$t/H^{\gamma}$\footnote{Figures 9--11 and 14 are reprinted from Physica B, vol~326, MacLaughlin D~E, Rose M~S, Young B~L, Bernal O~O, Heffner R~H, Morris G~D, Ishida K, Nieuwenhuys G~J and Sonier J~E, ``{$\mu$SR} and {NMR} in f-electron non-{Fermi}-liquid materials,''  pp~381--386, Copyright \copyright\ (2003), with permission from Elsevier.}. For CePtSi the best scaling is achieved with $\gamma = 1.6 \pm 0.1$. As before, this procedure does not require knowledge of the functional form of $\overline{G}(t,H)$. For CePtSi$_{0.9}$Ge$_{0.1}$ scaling is not obeyed for $H \lesssim 300$~Oe, but above this field scaling is found with the same value of $\gamma$. This value is considerably larger than in UCu$_{5-x}$Pd$_x$ for $T \lesssim 1$~K, where $\gamma(x{=}1.0) = 0.35$ and $\gamma(x{=}1.5) = 0.5$--0.7~\cite{MBHN01,MHBN02} (cf.\ \sref{sec:ucupd}).

\begin{figure}[ht]
\flushright \includegraphics[clip=,width=4in]{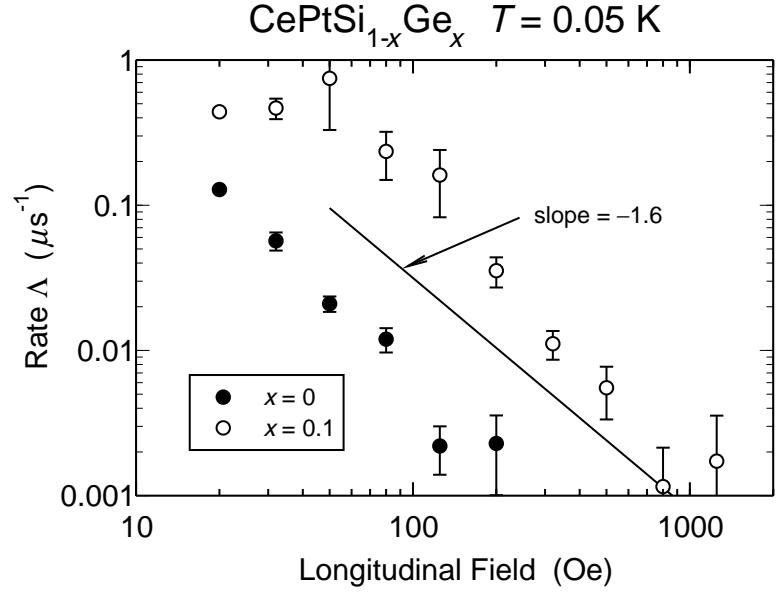}
\caption{Dependence of LF-$\mu$SR stretched-exponential relaxation rate~$\Lambda$ on longitudinal field in the NFL alloys CePtSi$_{1-x}$Ge$_x$, $x = 0$ and 0.1. From reference \protect\cite{MRYB03}.}
\label{fig:NFLfig3}
\end{figure}

The data are well described by the stretched exponential function~$A\exp[-(\Lambda t)^K]$ introduced in \sref{sec:muonrelax}, as shown in \fref{fig:NFLfig1} for CePtSi$_{0.9}$Ge$_{0.1}$. (For both Ge concentrations $K$ is constant at about 0.25 for fields above $\sim$50~Oe.) The field dependence of $\Lambda$ is plotted in \fref{fig:NFLfig3} 
for $T = 0.05$~K\@. For $x = 0$ $\Lambda$ varies as $H^{-\gamma}$, $\gamma \approx 1.6$\footnote{For $G(t,H) = G[\Lambda(H) t]$ time-field scaling~$G(t,H) = G(t/H^\gamma)$ implies $\Lambda(H) \propto H^{-\gamma}$.}, whereas for $x = 0.1$ this is true (with roughly the same exponent) only above a crossover field of about 100~Oe. This behaviour and the fact that $\gamma$ is greater than 1 are both consistent with a stretched-exponential spin autocorrelation function~$q(t) = \exp[-(t/\tau_{\rm c})^{\gamma-1}]$~\cite{KMCL96,KBCL01}, rather than the power-law~$q(t) = t^{1-\gamma}$ ($\gamma < 1$) found in UCu$_{5-x}$Pd$_x$. The crossover from $\Lambda \approx \rm const.$ to $\Lambda \propto H^{-\gamma}$ occurs for $\omega_\mu\tau_{\rm c} \sim 1$. The data of \fref{fig:NFLfig3} indicate that $\tau_{\rm c}$ is considerably longer for CePtSi than for CePtSi$_{0.9}$Ge$_{0.1}$, due perhaps to less pinning of fluctuations in the more ordered end compound. This difference in behaviour is related to details of the microscopic mechanism for the disordered spin dynamics, and is not well understood.

\subsection{YbRh\/$_2$Si\/$_2$ \rm\protect~\cite{IMBH03,IMOK03,IMOK03a}\label{sec:ybrhsi}}

The NFL compound~YbRh$_2$Si$_2$ seems to be a suitable system for the study of `clean' NFL physics. A number of bulk measurements~\cite{TGML00} indicate that YbRh$_2$Si$_2$ exhibits antiferromagnetism with a very low N\'eel temperature~$T_{\rm N} \approx 70$~mK, with a field-induced QCP at the orientation-dependent critical field~$H_{\rm c}$ that suppresses $T_{\rm N}$ to zero. Resistivity and specific heat measurements at low temperatures show $\Delta\rho \propto T$ and $C_{\rm el}/T \propto -\ln T$ over a temperature range of more than a decade. The NFL behaviour is suppressed and Fermi-liquid behaviour is recovered by the application of fields $> H_{\rm c}$. These results suggest that in zero field YbRh$_2$Si$_2$ is quite close to a QCP.

\begin{figure}[ht]
\flushright \includegraphics[clip=,width=4in]{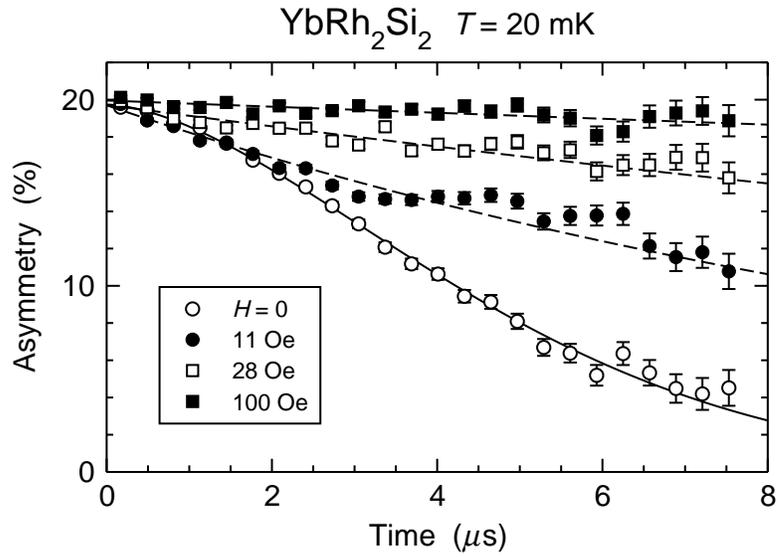}
\caption{LF-$\mu$SR relaxation in YbRh$_2$Si$_2$, $T = 20$~mK, longitudinal field~$H = 0$, 11, 28, and 100~Oe. Solid curve: fit to static Kubo-Toyabe relaxation function for $H = 0$. Dashed curves: fits to exponential relaxation function for $H > 0$. From reference \protect\cite{IMBH03}.}
\label{fig:YbRhSifig1}
\end{figure}

\Fref{fig:YbRhSifig1} 
gives the relaxation function~$\overline{G}(t,H)$ for the muon decay asymmetry in YbRh$_2$Si$_2$ at $T = 20$~mK~\cite{IMBH03}\footnote{Figures 12 and 13 are reprinted from Physica B, vol~326, Ishida K, MacLaughlin D~E, Bernal O~O, Heffner R~H, Nieuwenhuys G~J, Trovarelli O, Geibel C and Steglich F 2003, ``Spin dynamics in a structurally ordered non-{Fermi} liquid compound: {YbRh$_2$Si$_2$},'' pp~403--405, Copyright \copyright\ (2003), with permission from Elsevier.}\@. At zero field the data can be fit to a static Kubo-Toyabe (K-T) relaxation function~\cite{HUIN79} corresponding to a random distribution of static local fields with rms $\mathrm{width} \approx 2$~Oe. These static fields can be attributed to weak ordered Yb moments ($10^{-3}$--$10^{-2}~\mu_B$) in the antiferromagnetic state (N\'eel temperature~$T_N = 70$~mK), since the K-T rate is much larger than expected from nuclear dipolar fields and sets in only below $T_N$. The relaxation data for nonzero applied field can be fit to an exponential relaxation function~$\overline{G}(t,H) = \exp[-t/T_1(H)]$. An applied field of 11~Oe is more than five times larger than the estimated field at the muon site due to static Yb-moment magnetism~\cite{IMOK03a}, and therefore is large enough to decouple this static field. Thus the relaxation observed for $H \gtrsim 10$~Oe is dynamic. Therefore the field dependence of the rate~$1/T_1$ is not due to decoupling, but instead suggests a significant frequency dependence to the local-field fluctuation spectrum at the low muon frequencies. 

It is important to note that the observed exponential form for $H \ge 11$~Oe is evidence that the relaxation rate is substantially uniform throughout the sample, since as described above in \sref{sec:timefield} the signature of relaxation-rate inhomogeneity is a sub-exponential relaxation function, i.e., upward curvature of $\overline{G}(t)$ on a semi-log plot~\cite{MBHN01}. The muon spin relaxation is therefore probing spin fluctuations in a structurally ordered NFL system.

\begin{figure}[ht]
\flushright \includegraphics[clip=,width=4in]{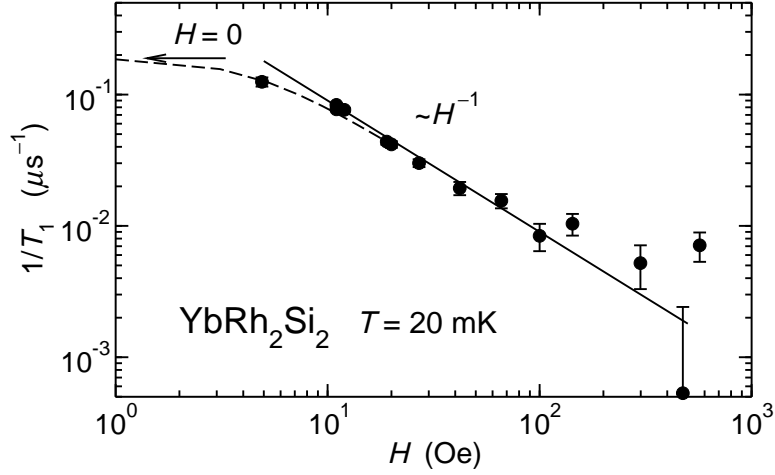}
\caption{Longitudinal field dependence of LF-$\mu$SR relaxation rate~$1/T_1$ in YbRh$_2$Si$_2$, $T = 20$~mK\@. Arrow: zero-field static K-T rate from fit of \protect\fref{fig:YbRhSifig1}. The muon decay lifetime limits accuracy for $1/T_1 \lesssim 0.01~\mu{\rm s}^{-1}$. From reference \protect\cite{IMBH03}.}
\label{fig:YbRhSifig2}
\end{figure}

\Fref{fig:YbRhSifig2} 
gives the dependence of $1/T_1$ on longitudinal field at $T = 20$~mK\@. The zero-field rate given by the arrow is the static K-T value obtained from the fit shown in \fref{fig:YbRhSifig1}. The relaxation rate shows a weak field dependence for magnetic fields less than $\sim$10~Oe but varies more strongly, as $H^{-\gamma}$, $\gamma = 1.0 \pm 0.1$, for higher fields. 

As discussed in \sref{sec:muonrelax}, this behaviour is consistent with a scaling law of the form~$\chi''(\omega,T) \propto \omega^{-\gamma} F(\omega/T)$, $F(x) \to x$ for small $x$. This scaling leaves $1/T_1(\omega,T)$ independent of temperature, in rough agreement with the observed weak ($\sim {-\ln T}$) dependence at $H = 19$~Oe (not shown; see reference~\cite{IMOK03}). Furthermore, for $\gamma \approx 1$ $q(t)$ varies slowly with $t$~\cite{KMCL96}; the low-temperature spin fluctuations are very long lived.

The exponential behaviour of the relaxation function in the present LF-$\mu$SR experiments indicates that YbRh$_2$Si$_2$ is an ordered stoichiometric compound. Thus disorder-driven theories~\cite{MDK96,CNJ00} seem to be ruled out for this material. The $\mu$SR results strongly suggest that YbRh$_2$Si$_2$ is a compound in which NFL behaviour is induced by homogeneous critical spin fluctuations.

\section{Effect of disorder on NFL behaviour in heavy-fermion metals} \label{sec:correl}

\begin{figure}[ht]
\flushright \includegraphics[clip=,width=4in]{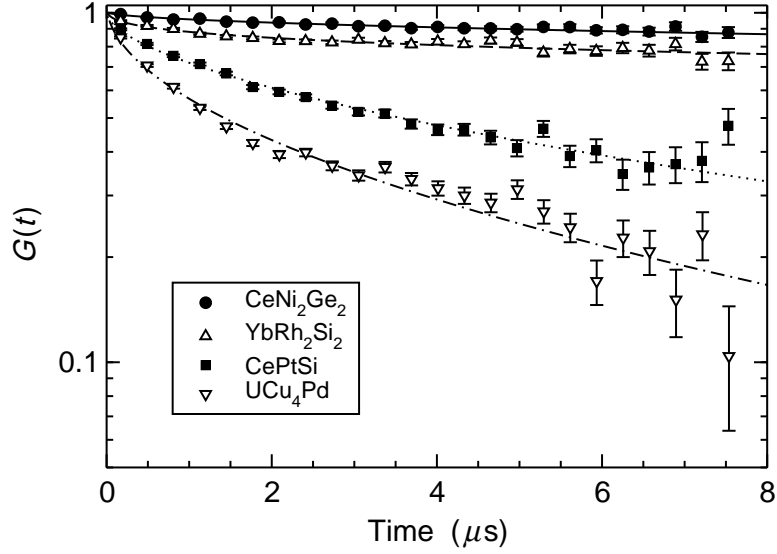}
\caption{LF-$\mu$SR relaxation functions~$\overline{G}(t,H)$ in NFL materials. (\fullcircle)~CeNi$_2$Ge$_2$, $T = 20$~mK, $H = 0$. (\opentriangle)~YbRh$_2$Si$_2$, $T = 20$~mK, $H = 32$~Oe. (\fullsquare)~CePtSi, $T = 0.05$~K, $H = 32$~Oe. (\opentriangledown)~UCu$_4$Pd: $T = 0.05$~K, $H = 20$~Oe. From reference \protect\cite{MRYB03}.}
\label{fig:NFLfig4}
\end{figure}

We next consider relations between LF-$\mu$SR relaxation and other properties in a number of NFL systems. \Fref{fig:NFLfig4} 
gives the muon spin relaxation functions in CeNi$_2$Ge$_2$, YbRh$_2$Si$_2$, CePtSi, and UCu$_4$Pd at low temperatures~\cite{MRYB03}. In these experiments a small longitudinal magnetic field has been applied, if necessary, to decouple any nuclear dipolar field, so that the relaxation is purely dynamic. The estimated relaxation functions for CeCu$_{5.9}$Au$_{0.1}$~\cite{AFGS95} and Ce(Ru$_{0.5}$Rh$_{0.5}$)$_2$Si$_2$~\cite{YMTK99,YMTO99} are comparable to that for CeNi$_2$Ge$_2$, i.e., the relaxation rates are also very low in these systems. CePtSi and UCu$_4$Pd show the most rapid and most nonexponential relaxation, whereas muons in CeCu$_{5.9}$Au$_{0.1}$, Ce(Ru$_{0.5}$Rh$_{0.5}$)$_2$Si$_2$, CeNi$_2$Ge$_2$, and YbRh$_2$Si$_2$ all relax much more slowly. YbRh$_2$Si$_2$ exhibits weak exponential muon spin relaxation at low temperatures and fields, with a rate~$1/T_1$ that varies inversely with longitudinal field~\cite{IMBH03} (cf.\ \sref{sec:ybrhsi}).

Thus there seems to be a range of relaxation behaviour in NFL materials. Is this due to differences in characteristic energy scales, or is disorder intimately involved?

\begin{figure}[ht]
\flushright \includegraphics[clip=,width=4in]{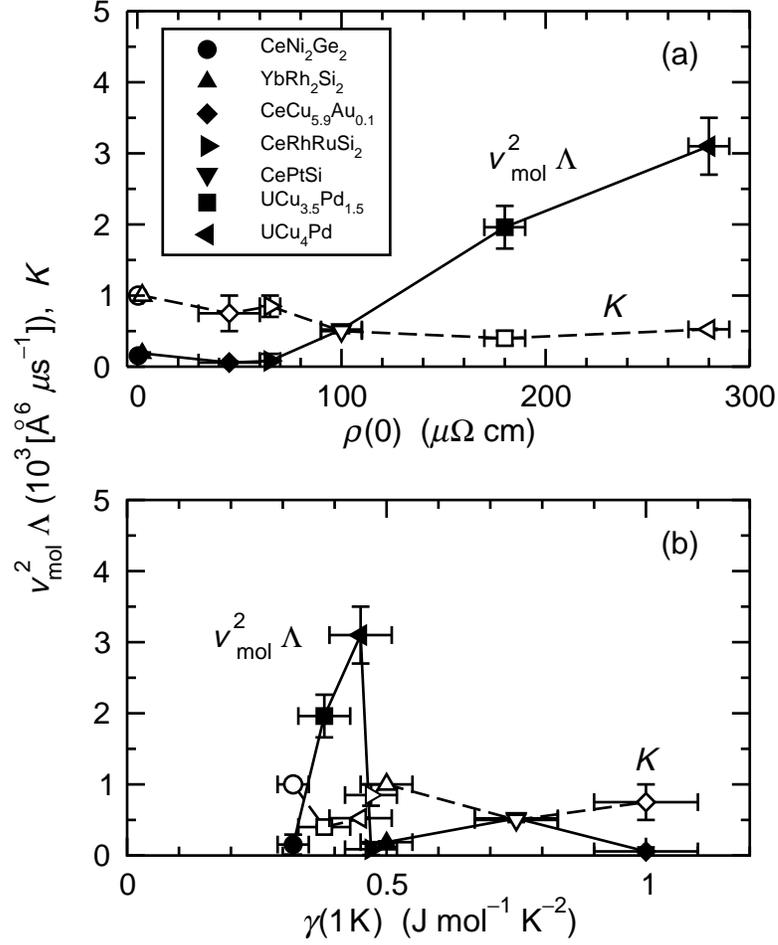}
\caption{Normalized muon stretched-exponential relaxation rate~$\Lambda$ (filled symbols) and exponent~$K$ (open symbols) vs.\ ($a$) residual resistivity~$\rho(0)$ and ($b$) specific heat coefficient~\mbox{$\gamma(T{=}1$\,K)}. Note that $\Lambda$ and $K$ do not have the same dimensions; they are plotted on the same graph only for convenience.}
\label{fig:correl}
\end{figure}

The low-temperature muon spin relaxation data for these materials have been fit to the stretched-exponential form~$\overline{G}(t,H) = \exp[-(\Lambda t)^K]$ as described above. Normalizing $\Lambda$ by $v_{\rm mol}^2$, where $v_{\rm mol}$ is the volume per mole of f ions, roughly accounts for the concentration of f-ion relaxing centres\footnote{The dipolar interaction between a muon and an f moment at distance~$r$ varies as $r^{-3} \propto v_{\rm mol}^{-1}$, as does the RKKY indirect interaction in simple models. This interaction enters the transition rate as the matrix element squared, so that the f-ion concentration dependence is normalized out after multiplication by $v_{\rm mol}^2$. Our conclusions are not sensitive to this normalization.}. \Fref{fig:correl}\,($a$)
gives plots of $v_{\rm mol}^2\Lambda$ and $K$ vs.\ the residual resistivity~$\rho(0)$~\cite{MRYB03,MacL03}. A smooth and marked increase of $v_{\rm mol}^2\Lambda$ with increasing $\rho(0)$ is observed, together with a decrease of $K$ from $\sim$1 to $\sim$1/2. There appears to be a crossover from ordered to disordered behaviour for $\rho(0) \gtrsim 100~\mu\Omega$~cm; disorder-driven effects are small even in solid solutions such as CeCu$_{5.9}$Au$_{0.1}$ [$\rho(0) \sim 30~\mu\Omega$~cm]. This suggests that disorder is important in, and could even be the driving mechanism for, slow f-electron NFL spin dynamics.

We also wish to examine the question of correlation between the relaxation behaviour and some measure of the characteristic fluctuation rate, or characteristic energy scale, of the system. By definition NFL metals at low temperatures exhibit no characteristic energy other than the temperature. Nevertheless one can use a quantity such as the low-temperature Sommerfeld specific-heat coefficient~$\gamma(T) = C(T)/T$ (not to be confused with the time-field scaling exponent) as a rough gauge of the expected fluctuation rate, and hence of muon spin relaxation behaviour if slow fluctuations are correlated with large values of $\gamma(T)$. In NFL materials $\gamma(T)$ is temperature dependent, and we have arbitrarily chosen to use the value~\mbox{$\gamma(T{=}1\,\rm K)$}\@. The conclusions do not depend critically on this choice.

In \fref{fig:correl}\,($b$) the relaxation parameters are plotted against \mbox{$\gamma(T{=}1\,\rm K)$}, which is admittedly only a crude estimator of energy scales in these systems. Nevertheless, in a conventional picture smaller values of \mbox{$\gamma(T{=}1\,\rm K)$} would imply larger characteristic energies, faster spin fluctuations, more motional narrowing, and slower muon spin relaxation rates. This is not observed; 
there is no apparent correlation between muon spin relaxation parameters and \mbox{$\gamma(T{=}1\,\rm K)$}. It appears that disorder slows down the spin fluctuations, much as in the paramagnetic state of spin glasses but with no spin freezing down to the lowest temperatures of measurement ($\sim$20~mK)\@. We appear to be dealing with spin glass behaviour in the extreme quantum limit~\cite{GrRo99,RSBC01}.

\section{Concluding remarks}
\label{sec:discus}

\subsection{ Local or cooperative dynamics?} \label{sec:local?}

The muon spin relaxation results in disordered systems can be compared with existing local (single-ion or cluster) theories of disorder-driven NFL behaviour. The Kondo-disorder model cannot account for the order of magnitude of the experimental muon spin relaxation rates at low temperatures~\cite{MacL00}; there simply are not enough slowly-fluctuating uncompensated moments at low temperatures to account for the observed muon relaxation rates. The situation for the Griffiths-McCoy singularity model is more complicated, however.

In an early version of the theory~\cite{CNCJ98}, which treats the effect of f-moment clustering, there is no dissipation in the f-electron spin dynamics, and the local cluster dynamic susceptibility is sharply resonant at a distributed characteristic tunneling energy~$E$:
\begin{equation}
\chi''(\omega,E) \propto \delta(\omega - E)\,\tanh(E/2T)\,.
\label{eq:reschi}
\end{equation}
Note that this form does not exhibit frequency scaling $\chi''(\omega) \propto \omega^{-\gamma}F(\omega/T)$. Together with the distribution function~$P(E) \propto E^{-1+\lambda}$, where $\lambda < 1$ is a nonuniversal scaling exponent, we have
\begin{equation}
\overline{\chi''}(\omega) = \int dE\, P(E)\chi''(\omega,E) \propto 
\omega^{-1+\lambda}
\tanh(\omega/2T) \,.
\label{eq:Grif}
\end{equation} 
This scaling form is in agreement with the INS data for $\lambda \approx 0.7$ and is reminiscent of the LF-$\mu$SR results. 

But in LF-$\mu$SR experiments it is the muon spin relaxation function~$G(t,H)$, not $\chi''(\omega)$, that is spatially averaged. As discussed in \sref{sec:timefield}, the observed time-field scaling of the spatially-averaged muon spin relaxation function~$\overline{G}(t,H)$ shows that the {\em local\/} $\chi''(\omega)$, not just the spatial average $\overline{\chi''}(\omega)$, scales as $\omega^{-\gamma}$. Local frequency scaling is not a property of the Griffiths-McCoy singularity model, in which the scaling is found only after the average has been taken, and only for a sharply resonant (and nonscaling) form [equation \eref{eq:reschi}] of the local $\chi''(\omega,E)$. A later form of this theory~\cite{CNJ00} considers dissipative effects, which broaden $\chi''(\omega,E)$ but do not give it a scaling form. Furthermore, if the width of $\chi''(\omega)$ is much greater than $\omega$ it is not hard to show that $\overline{\chi''}(\omega)$ no longer follows $P(\omega{=}E)$, so that this mechanism for scaling of $\overline{\chi''}(\omega)$ is lost. It is also difficult to see how the dynamic susceptibility of \cite{CNJ00} would yield the observed temperature dependence of $\Lambda(\omega_\mu,T)$. Thus the Griffiths-McCoy singularity theory does not seem to account for the $\mu$SR results for a number of reasons.

More generally, disorder-driven pictures that ascribes NFL behaviour to `rare' objects (low-$T_{\rm K}$ spins or large clusters) lead to spatial distributions of spin fluctuation rates (slow for the rare objects), and thus are not in agreement with time-field scaling. We conclude that local disorder-driven models account for inhomogeneous distributions of static susceptibility in disordered NFL materials (\sref{sec:ddNFL}), but that the distributed local dynamics in these models do not describe the observed muon spin relaxation. Time-field scaling of the relaxation data is strong evidence for a cooperative mechanism, i.e., spin dynamics with a single correlation function for the entire sample.

\subsection{Glassy behaviour?}

The question of local vs.\ cooperative dynamics seems to be answered in favor of the latter as discussed above, at least for the UCu$_{5-x}$Pd$_x$ system which has been studied most intensively. In contrast, a number of more general (and related) questions remain unanswered, or at best only partially answered:
\begin{itemize}

\item Disorder slows down the (cooperative) spin fluctuations in disordered NFL systems [\fref{fig:correl}\,($a$)]. Are these slow `glassy' fluctuations ordered-system quantum critical fluctuations {\em strongly modified\/} by disorder, or new `quantum glass' excitations {\em created\/} by disorder? Is there even a distinction in principle between these notions?

\item The correlation shown in \fref{fig:correl}\,($a$) strongly suggests that glassy dynamics depend `universally' on residual resistivity. What does this mean physically? Is there a picture in which disorder-driven slowing down of fluctuations in a metal depends only on the electronic mean free path? In this regard the increase of impurity scattering near a QCP predicted by Varma, Miyake, and co-workers~\cite{Varm97,MiNa02,MiMa02} may be relevant if there is a slowing effect of this scattering on spin lifetime in the system near criticality.

\item More generally, how are the fluctuations slowed down? Is {\em frustration\/} involved, in the sense that it is involved in slow spin fluctuations in spin glasses and geometrically frustrated systems?

\item Is there a mixing of (quantum) critical and Griffiths-McCoy singular behaviour? Do we need to invoke aspects of both phenomena~\cite{MMS02}?


\end{itemize}
The static and dynamic response of spins in f-electron NFL materials exhibits very strong effects of disorder, revealing poorly-understood instabilities of quantum criticality in these systems to strong inhomogeneity and glassy behaviour. It is to be hoped that the present experimental results lead to further endeavours in this area.

\ack{We wish to thank B~Andraka, R~Chau, C~Geibel, K~Heuser, W~Higemoto, R~Kadono, G~Luke, G~Morris, M~Rose, J~Sarrao, F~Steglich, J~Thompson, O~Trovarelli, Y~Uemura, and B-L Young for their collaboration on the $\mu$SR experiments reviewed in this paper, and A~Amato, D~Arsenau, C~Baines, M~Good, D~Herlach, B~Hitti, and S~Kreitzman for their invaluable help with these experiments. We have benefited greatly from conversations with M~Aronson, W~Beyermann, A~Castro Neto, P~Coleman, V~Dobrosavljevi{\'c}, Z~Fisk, K~Ingersent, G~Kotliar, A~Millis, E~Miranda, R~Osborn, L~Pryadko, Q~Si, and R~Walstedt. We are especially grateful to C~Varma and H~Maebashi for careful reading of the manuscript. Research described in this article was performed in part at the Swiss Muon Source, Paul Scherrer Institute, Villigen, Switzerland, and in part at TRIUMF, Vancouver, Canada. It was supported in part by the U.S. NSF, Grant nos.~0102293 (Riverside), 0203524 (Los Angeles) and 9705454 (San Diego), the U.S. DOE, Contract no.~DE-FG05-96ER45268 (Gainesville), the COE Research Grant-in-Aid (10CE2004) for Scientific Research from the Ministry of Education, Sport, Science, and Technology of Japan (Kyoto), the Canadian NSERC (Burnaby), and the Netherlands NWO and FOM (Leiden), and was performed in part under the auspices of the DOE (Los Alamos).}

\section*{References}



\end{document}